Brief Review

# Antimatter gravity and the Universe

Dragan Slavkov Hajdukovic
Institute of Physics, Astrophysics and Cosmology
Crnogorskih junaka 172, Cetinje, Montenegro
dragan.hajdukovic@cern.ch

**Abstract:** The aim of this brief review is twofold. First, we give an overview of the unprecedented experimental efforts to measure the gravitational acceleration of antimatter; with antihydrogen, in three competing experiments at CERN (AEGIS, ALPHA and GBAR), and with muonium and positronium in other laboratories in the world. Second, we present the 21$^{st}$ Century's attempts to develop a new model of the Universe with the assumed gravitational repulsion between matter and antimatter; so far, three *radically different* and *incompatible* theoretical paradigms have been proposed. Two of these 3 models, Dirac-Milne Cosmology (that incorporates CPT violation) and the Lattice Universe (based on CPT symmetry), assume a *symmetric Universe* composed of equal amounts of matter and antimatter, with antimatter somehow "hidden" in cosmic voids; this hypothesis produced encouraging preliminary results. The heart of the third model is the hypothesis that *quantum vacuum fluctuations are virtual gravitational dipoles*; for the first time, this hypothesis makes possible and inevitable to *include the quantum vacuum as a source of gravity*. Standard Model matter is considered as the only content of the Universe, while phenomena usually attributed to dark matter and dark energy are explained as the local and global effects of the gravitational polarization of the quantum vacuum by the immersed baryonic matter. An additional feature is that we might live in a cyclic Universe alternatively dominated by matter and antimatter. In about three years, we will know if there is gravitational repulsion between matter and antimatter; a discovery that can forever change our understanding of the Universe.

## 1. Prelude

Nine decades after the discovery of antimatter, we don't know the answer to the simplest question: In which direction would an anti-apple fall in the gravitational field of the Earth, down or up? We all know that an apple falls down, but, no one knows if an anti-apple would also fall down *or would fall up*. The aim of this review is: (1) a *basic* description of experiments that would answer this fundamental question and (2) to provide an elementary understanding of theoretical speculations about the possible impact (in astrophysics and cosmology) of eventual affirmation of gravitational repulsion between matter and antimatter.

During the first few decades after the discovery of antimatter (roughly 1930-1960), which are also the first decades of modern cosmology, gravitational repulsion between matter and antimatter was considered as a serious possibility. However, everything changed during the 7$^{th}$ decade of 20$^{th}$ Century, when, further thinking about a negative gravitational charge (gravitational mass) of antimatter was *suppressed by purely theoretical arguments* that antimatter must fall in the same way as matter. In brief, *despite the absence of experimental evidence*, the gravitational attraction between matter and antimatter was imposed as an absolute truth; any questioning of this prescribed truth was highly damaging to the scientific reputation of rare scientists that had the courage to oppose group-thinking. It is fair to say that the authors of arguments, great scientists and deep thinkers, are not responsible for the later dark period of suppression of alternative thinking. Our understanding of antimatter gravity in the 20$^{th}$ Century is well reviewed[1] in the article "The arguments against "antigravity" and the gravitational acceleration of antimatter".

The most beautiful and ingenious of 3 major arguments against "antigravity" was given by Good[2]; roughly speaking, if "antigravity" exists, the stationary states of a neutral kaon system (composed of a quark and an antiquark) would be perturbed with an inevitable violation of CP symmetry. It was before the discovery of CP violation when a huge majority of physicists (too huge to be right!) was convinced that CP is an exact symmetry of nature; consequently, the presumed non-existence of CP violation was taken as an argument against gravitational repulsion between matter and antimatter.

Perhaps, the most serious arguments against "antigravity" are improved versions (for a brief overview see Ref. 3, Section 5.5) of the initial Schiff's argument[4]. The essence of these arguments is that *virtual particle-antiparticle pairs* (i.e. the quantum vacuum) dominate the mass of nucleons and hence atoms. Consequently, gravitational charges of the opposite sign should produce observable violation of the universality of free fall of macroscopic bodies made of different materials. Of course, this argument seems very plausible. However, even if we neglect already known shortcomings of Schiff's calculations[1], and recent contra-arguments[5], my question is why we should trust this theoretical prediction after the theoretical debacle called the cosmological constant problem[6]; the essence of the cosmological constant problem is that the gravitational impact of the quantum vacuum is at least forty orders of magnitude larger than permitted by empirical evidence. How we can trust our calculation of the gravitational impact of the quantum vacuum in one case if it so dramatically wrong in the other case? Let us note (thank you to the excellent Reviewer 1 for this remark) that this argument is not critical in theories that do not use quantum vacuum as a source of gravity (i.e. Dirac-Milne Cosmology and the Lattice Universe, presented in Section 3.1).

In any case this review is not about already reviewed[1,5] theoretical arguments. This review is exclusively about antimatter gravity experiments and cosmological theories that start with the hypothesis of surprising outcomes of these experiments (i.e. gravitational repulsion between matter and antimatter).

Before we continue, let us clarify the relation between the fundamental CPT symmetry and the gravitational repulsion between matter and antimatter. CPT is violated by gravity *only if* antimatter-antimatter interaction *is different* from matter-matter interaction. Pictorially speaking, according to CPT, an anti-apple must fall in the gravitational field of an anti-Earth in the same way as an apple falls in the gravitational field of the Earth. However, the predictive power of CPT is not enough to impose constraints on matter-antimatter interaction; from the point of view of CPT, both the gravitational attraction and the gravitational repulsion between matter and antimatter are possible.

In principle, a theory of gravity can be compatible, but also can be incompatible with CPT symmetry; at the present stage of our knowledge CPT is not a criterum of validity of a theory of gravity. An intriguing and plausible argument was given by Villata[7] that General Relativity and CPT are compatible only if there is gravitational repulsion between matter and antimatter.

After suppression during the 20$^{th}$ Century, in the beginning of 21$^{st}$ Century we witness a strong, both experimental[8-15] and theoretical[16-29], "rebellion" of the hypothesis of gravitational repulsion between matter and antimatter. This rebellion will end within the next 3-4 years; the outcome of this rebellion will be either an experimental disproval of the hypothesis or *a scientific revolution*.

We live in a time of unprecedented experimental efforts to measure the gravitational acceleration of antimatter. After more than one decade of complex development, three competing experiments at CERN (ALPHA-g[8,9], AEGIS[10] and GBAR[11]) are close to the measurement of the gravitational acceleration of antihydrogen. However, because of the shut-down of CERN that will last for the next 2 years, the first measurements must be postponed until the end of 2021 or the beginning of 2022. In addition to experiments with antihydrogen at CERN, different experimental groups in other laboratories are preparing experiments with muonium[12] and positronium[13]. Of course, complete empirical evidence must be the result of two complementary efforts: experiments in laboratories and astronomical observations. It is encouraging that there are already different ideas for astronomical observations; for instance, a study of orbits of tiny satellites in trans-Neptunian binaries has the potential to reveal the eventual gravitational impact of the quantum vacuum[14,15]. While a huge majority of theoretical physicists (perhaps too huge to be right) believe that *the result of experiments is known in advance*, i.e. that antimatter falls exactly in the same way as matter, it may be a good idea to wait and see.

In parallel with the inevitably long preparation of extremely difficult and sophisticated experiments, different authors have tried to reveal the eventual astrophysical and cosmological consequences of repulsive gravity between matter and antimatter. So far, three *radically different* and *completely incompatible* theoretical paradigms have been proposed.

Two of these 3 models (Dirac-Milne Cosmology[16-18] and the Lattice Universe[19-21]) assume a *symmetric Universe* composed of *equal amounts of matter and antimatter*, with antimatter somehow "hidden" in cosmic voids. However, as we will see below, after this common assumption these two models diverge. Dirac-Milne Cosmology introduces CPT violation while within the Lattice Universe CPT Symmetry is respected; in Dirac-Milne Cosmology the Universe expands with constant speed *c* while in the Lattice Universe, expansion of the Universe is accelerated. In both models' preliminary results are intriguing and encouraging.

At the heart of the third model[22-29] is the working hypothesis that quantum vacuum fluctuations are virtual gravitational dipoles; for the first time, this hypothesis makes possible and inevitable to *include the quantum vacuum as a source of gravity*. The Standard Model matter (i.e. matter made of quark and leptons interacting through the exchange of gauge bosons) is considered as the only content of the Universe, while phenomena usually attributed to dark matter and dark energy are explained as the local[23,26,29] and global[27,29] effects of the gravitational polarization of the quantum vacuum by the immersed baryonic matter. An additional feature is that we might live in a cyclic Universe[22,24,29] with cycles alternatively dominated by matter and antimatter; we live in a Universe dominated by matter because the previous cycle was dominated by antimatter.

This variety of models that exclude each other is welcome in a period of great crisis in physics when we are trying to guess how Nature works.

While this Review is strongly limited to *antimatter gravity* it is important to mention a remarkable series of papers[30-34], which are completely outside of the field of antimatter gravity research. The key point is that, in order to reconcile MOND and Dark Matter paradigms, a *negative* gravitational charge (let us underscore again, *completely unrelated to antimatter*) was introduced. The essence of this significant work is well described in the abstract of the initial paper[30]: "The modified Newtonian dynamics (MOND) has been proposed as an alternative to the dark matter paradigm; the philosophy behind is that there is no dark matter and we witness a violation of the Newtonian law of dynamics. In this paper, we interpret the phenomenology sustaining MOND differently, as resulting from an effect of 'gravitational polarization', of some cosmic fluid made of dipole moments, aligned in the gravitational field, and representing a new form of dark matter. We invoke an internal force, of non-gravitational origin, in order to hold together the microscopic constituents of the dipole. The dipolar particles are weakly influenced by the distribution of ordinary matter; they are accelerated not by the gravitational field, but by its gradient or tidal gravitational field." Hence, the hypothetical "*dipolar dark matter*" is composed of *permanent* gravitational dipoles (i.e. gravitational charges of the opposite sign); it is obvious that the existence of dipoles of any kind (electric, magnetic, gravitational…) assures the existence of the corresponding polarization (i.e. to some extent, dipoles are aligned in an external field). For completeness, let us note the useful calculations[35,36] that are similar to the "dipolar dark matter" paradigm, but without a specified nature of gravitational dipoles.

It is intriguing and encouraging that very different theoretical motivations and approaches [reconciliation of MOND and Dark Matter paradigms, compatibility[7] of General Relativity and CPT symmetry; the simplest possible solution to the cosmological constant problem (See Section 3.2)] lead in the same direction: the existence of negative gravitational charge.

## 2. Antimatter Gravity Experiments

The beginning of the 21st century is marked by

- Three, that have been on-going for more than a decade, active experiments at CERN (ALPHA-g[8,9], AEGIS[10] and GBAR[11]) which are competing to be the first to measure the gravitational acceleration of antihydrogen.
- Preparation of analogous experiments with muonium[12] (an exotic atom made of an antimuon and an electron) and positronium[13] (a system composed of an electron and an antielectron).

An amusing (while non-scientific) question is why so many great experimentalists waste significant time on experiments whose outcomes are *known in advance* (according to the nearly unanimous prediction of theorists' antimatter must fall in the same way as matter).

All experiments described below will measure the impact of the gravitational field of the Earth on antimatter. Hence, we will get the first empirical evidence about gravitational interaction between *matter and antimatter*, while the gravitational interaction between *antimatter and antimatter* will garner no empirical evidence whatsoever.

## 2.1 ALPHA-g Experiment

ALPHA is an extremely successful if not the best antimatter experiment of all time. In 2010, the ALPHA collaboration achieved the first-ever trapping of cold antihydrogen atoms; a seminal success, opening a new era in the study of antimatter. From that time on, for the ALPHA team, production and trapping of antiatoms has become routine, making possible a long-awaited spectroscopy of antihydrogen as a fundamental tool to look for the eventual differences between matter and antimatter.

After two general purpose traps (ALPHA-1 and ALPHA-2) the ALPHA Collaboration has recently constructed a new ALPHA-g apparatus[9] devoted to the measurement of the gravitational acceleration of antihydrogen. The experiment is pragmatically divided in two stages; in the first stage the goal is limited to the much easier task of determining *the sign of the acceleration*, while the precise measurement of acceleration is left for the second stage. It is important and encouraging that a proof-of-principle measurement[8] has already been completed.

In brief (See Fig. 1), ALPHA-g consists of a vertically oriented apparatus consisting of two symmetric atom and Penning trap arrangements with a high precision region in the centre. Surrounding the cryostat is a radial time projection chamber tracking detector used for locating antiproton annihilations within the trapping volume. The symmetry of the design is aimed towards conducting equivalent experiments on either end of the trap to set limits on or cancel systematic construction and detection errors.

An important feature is that the magnetic potential in the vertical direction (y-direction in Fig.1) can be tuned via independent control of the trap mirror currents.

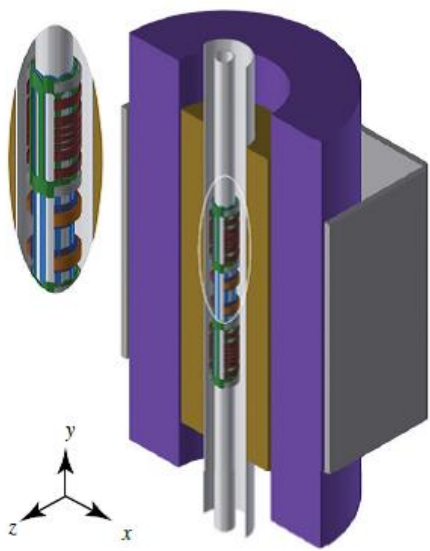

**Figure 1**. Schematic of the ALPHA-g magnet system, with its cylindrical axis of symmetry oriented in the vertical (y) direction. An external solenoid (purple) generates the uniform solenoidal field required for internal Penning traps and operation of the radial time projection chamber detector (gold). Inset shows details on the upper and precision trap. Two independent atom traps surrounding Penning traps are generated by a set of seven mirror coils (red) and a short octupole (green). A precise analysis trap is formed between the two dark orange coils and a long octupole (blue). Adiabatic transport of antihydrogen atoms between trapping can be accomplished through sequencing four transfer mirror coils (grey). External magnetic error fields can be corrected through rectangular correction coil panels (dark grey). Possible trapping regions range in length from approximately 280 mm (single end atom trap) up to 1.3 m (between the extremes of the two end traps). From Reference 9.

The ALPHA-g trap depth is approximately $540 mK$ for antihydrogen atoms born at the centre of the magnetic volume (i.e. only atoms with energy smaller than $540 mK$ are trapped). For sub-$540 mK$ antihydrogen in a trap roughly 280mm tall, atoms will bounce over the height approximately 1000 times in 10 seconds. The gravitational potential difference for hydrogen over a distance of approximately 280 mm (i.e. during one bounce) and the corresponding magnetic potential change are roughly equal. After the opening of the trap one fraction of bouncing atoms will be detected at the bottom and the other at the top. Based on the measured values of these fractions, the sign of the acceleration can be determined; for it, a few hundred antihydrogen annihilation events are needed—a data rate which is presently achievable during a single 8 h shift on ALPHA.

## 2.2 AEGIS Experiment

The competing AEGIS experiment[10] plans to measure the vertical deviation of a pulsed horizontal beam of cold antihydrogen atoms; the vertical deviation, which is expected to be a few microns, would be measured using a Moire deflectometer. This is visualised in Fig. 2.

A horizontal beam of antiprotons enters the "moiré" setup consisting of three equally spaced elements: two gratings and a spatially resolving emulsion detector. The two gratings with periodicity $d$ define the classical trajectories leading to a fringe pattern with the same periodicity at the position of the detector.

In the transit time of the particles through the device is known, absolute force (and the corresponding acceleration) measurements are possible by employing Newton's second law of mechanics.

To infer the force, the shifted position of the "moiré" pattern must be compared with the expected pattern without force. This is achieved using light and near-field interference, the shift of which is negligible.

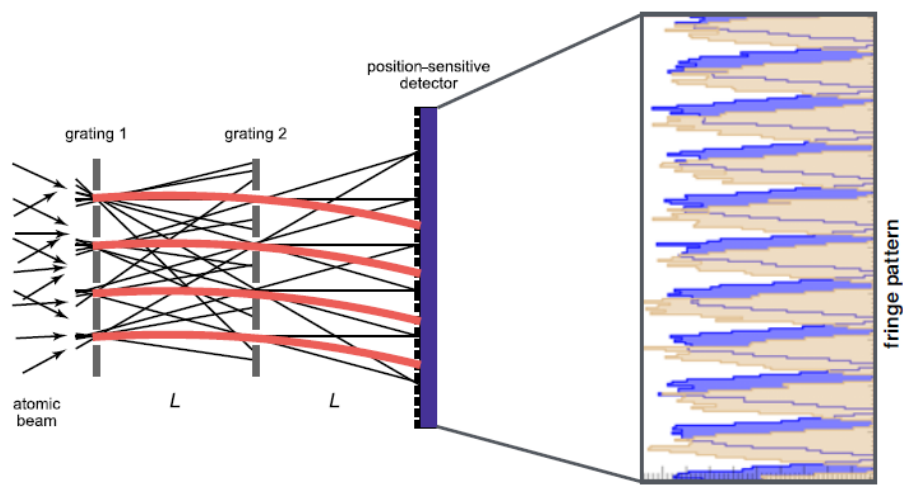

**Figure 2**. Illustration of the Moiré deflectometer technique used in AEGIS experiment. A divergent antihydrogen beam propagates through two identical gratings. Antihydrogen atoms passing the gratings follow a parabolic path and annihilate on a position sensitive detector; the annihilation points form a fringe pattern which is shifted in the presence of a force (it would be shifted down in the case of gravitational attraction and shifted up in the case of the gravitational repulsion). From Reference 10.

## 2.3 GBAR Experiment

The third competing experiment GBAR[11] is the closest one to our classical vision of a free-fall experiment: the measurement of the time of flight corresponding to a known change of the height.

The fixed change of the height in the GBAR experiment would be small (roughly about $20 cm$) and for the measurement to be successful the initial speed of $\bar{H}$ atoms mustn't be bigger than a few metres per second; a speed about 3-4 orders of magnitude smaller than the speed of antihydogen in the ALPHA and AEGIS experiments. Hence, while the final measurement in GBAR is more direct and simpler than in the competing experiments, the preparation of the needed ultra-cold antihydrogen is more difficult task and requires to this point the unprecedented cooling of antihydrogen. Just to get an idea about complexity common to all antimatter gravity experiments let us give a few more details.

In the first step GBAR would not produce atoms of antihydrogen (antiproton and positron) but rather *antihydrogen ions* (antiproton with *two* positrons). This is motivated by the fact that ion-cooling techniques are more efficient than techniques of cooling neutral atoms.

In the second step antihydrogen ions would be sympathetically cooled with laser cooled matter ions such as $Be^+$ to temperatures of less than $10\mu K$ (i.e. with velocities of the order of $0.5\ m/s$). After that, the extra positron may be photo detached by a laser pulse, with energy of only a few $\mu eV$ above the threshold, in order to obtain an ultracold antiatom. The time of flight of the resulting free fall should be about $200ms$, which can be easily measured to extract the acceleration due to Earth's gravity.

### 2.4 Experiments with positronium

Positronium (Ps) is a hydrogen-like atom composed of an electron and a positron. The ground state lifetime of triplet Ps is $1.4 \times 10^{-7}s$, while for the measurement of gravitational acceleration, lifetimes of a few milliseconds or greater are required. Fortunately, the lifetime of positronium is an increasing function of the principal quantum number $n$; intuitively it can be understood as the decrease of annihilation rate because larger $n$ means a larger distance between the electron and the positron. For a given $n$, the lifetime is longer for the higher values of the angular momentum; the lifetime increases[3] as $n^3$ for non-circular states and $n^5$ for circular states (i.e. states with maximal angular momentum $l = |m| = n - 1$). For instance, lifetime corresponding to $n = 30$ is respectively a few milliseconds and a few seconds for non-circular and circular states.

Hence, the gravitational experiments are possible only with positronium atoms optically excited to long-lived Rydberg states (i.e. states with large $n$). The good news is that there is an encouraging initial success in the creation of excited states of positronium by laser.

An experimental programme[13] currently underway at University College London (UCL) has as its *long-term goal* a gravitational free-fall measurement of positronium atoms. On their long way to success they must overcome many obstacles; among these are the production of positronium atoms in a cryogenic environment, efficient excitation of these atoms to suitably long-lived Rydberg states, and their subsequent control via the interaction of their large electric dipole moments with inhomogeneous electric fields.

Let us underscore that the experiment with the antihydrogen is an experiment in the quark sector, while the experiment with positronium would be in the lepton sector of the Standard Model.

### 2.5 Experiments with muonium

Muonium (Mu) is a hydrogen-like system composed of an antimuon $\mu^+$ (which is unstable with a lifetime equal to $2.2\mu s$) and an electron $e^-$. Measuring muonium gravity — if feasible — would be the *first* gravitational measurement of a 2nd-generation particle of the Standard Model.

It is obvious that the lifetime of muonium is limited to the lifetime of antimuon and cannot be made longer. Despite the obstacle of a very short lifetime (and some other obstacles) "The Muonium Antimatter Gravity Experiment (MAGE)" is in development[12].

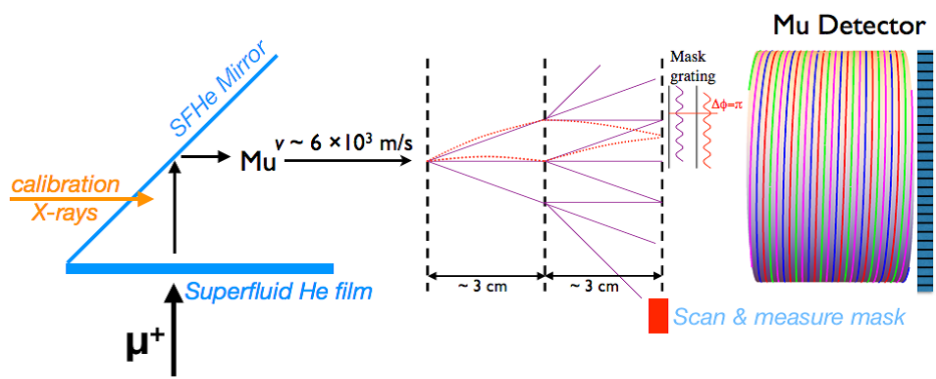

**Figure 3.** Schema of the MAGE experiment. From Reference 12.

The MAGE, which is in fact a difficult application of well-established atom interferometry, is illustrated in Fig. 3. A horizontal muonium beam is directed into a three-grating interferometer. The first two gratings set up an interference pattern that has the same period as the gratings. Gravity causes a phase shift in the interference pattern which is equivalent to the deflection of an individual muonium atom; this phase shift (and hence the gravitational acceleration) is determined by scanning the third grating vertically by several grating periods and measuring the resulting sinusoidal variation in detected Mu intensity. The interferometer is aligned using a soft X-ray beam with wavelength comparable to that of the Mu. This X-ray beam will also be used to determine the phase of an undeflected beam.

Just to get an idea of the complexity of this (and all other) antimatter gravity experiments please note that the Schema of the MAGE experiment contains a "*Superfluid He film*". Let's explain why.

Antimuons $\mu^+$ do not annihilate on contact with ordinary matter (since there are no $\mu^-$ in ordinary matter). Once stopped, $\mu^+$ can combine with a free electron to form cold muonium. Such production of muonium was achieved in many different materials. The trouble is that muonium atoms are emitted with a thermal velocity distribution and uniformly in $\cos \alpha$ ($\alpha$ being the angle relative to the surface). With such a source of muonium the gravity experiment is impossible. In order to perform this gravity experiment we must have a parallel monochromatic beam of muonium; otherwise the interference patterns of different atoms will have different phases. At this point the superfluid $He$ film is a solution. In fact, muonium has a negative chemical affinity in superfluid helium, so when it diffuses out of the liquid, it is ejected with a nearly *constant* velocity ($6300\,m/s$) *normal* to the surface. Hence, a superfluid He film is an ideal source of muonium, a parallel nearly monochromatic beam ($\Delta E/E \approx 0.2\%$).

## 3. Theoretical speculations

What if experiments establish that antimatter falls up?

We must respect two experimental facts. First, particles and antiparticles have the same inertial mass ($m_i = \bar{m}_i$, bar denotes an antiparticle). Second, according to the Weak Equivalence Principle (WEP), which is based on indisputable empirical evidence, for each body *made of matter*, we can use the equality $m_i = m_g$, where $m_g$ denotes the gravitational mass (it may be better to say the gravitational charge) of matter that is used in Newton's law of gravity. In principle, we must distinguish between active gravitational mass $m_{ga}$ (as a source of the gravitational field) and passive gravitational mass $m_{gp}$ (as a measure of the gravitational force acting on a body in a given gravitational field); fortunately for matter $m_g = m_{ga} = m_{gp}$ and we are free of the complication of two different gravitational charges.

Strictly speaking, if, in the gravitational field of the Earth, antimatter falls up, within the framework of Newtonian gravity it means that antiparticles have a *negative* passive gravitational charge. We cannot measure the active gravitational charge of antimatter, but the most plausible assumption (and the only assumption that respects CPT symmetry) is that, as in the case of matter, active and passive gravitational charges are equal ($\bar{m}_g = \bar{m}_{ga} = \bar{m}_{gp}$), while the mass and gravitational charge of an antiparticle have the opposite sign, i.e. $\bar{m}_g = -\bar{m}_i$ or more impressively $m_g + \bar{m}_g = 0$. Of course, we must stay open-minded; Nature may surprise us with matter-matter, matter-antimatter and antimatter-antimatter gravitational interactions that cannot be described by any combination of signs of passive and active gravitational mass[18].

Now, let us remember the cornerstones of contemporary Cosmology.

First, the cosmological principle, i.e. the statement that at any particular time the Universe is isotropic about every point (note that this includes homogeneity), leads to the Friedman-Lemaitre-Robertson-Walker (FLRW) metric:

$$ds^2 = c^2 dt^2 - R^2(t) \left[\frac{dr^2}{1-kr^2} + r^2(d\theta^2 + \sin^2\theta\, d\vartheta^2)\right], \qquad (1)$$

where k=+1, k=-1 and k=0 correspond respectively to a closed, open and flat Universe. The dynamics of the above space-time geometry is completely characterized by the scale factor R(t), which can be determined only within the framework of a specific theory of gravity.

Within the framework of General Relativity, the scale factor R(t) is the solution of Einstein's equation $G_{\mu\nu} = -(8\pi G/c^4)T_{\mu\nu}$. Einstein tensor $G_{\mu\nu}$ is determined by the FLRW metric, while the Energy-momentum tensor $T_{\mu\nu}$ is approximated by the energy-momentum tensor of a perfect fluid; characterised at each point by its proper density $\rho$ and pressure $p$.

If a cosmological fluid consists of several distinct components denoted by $n$, the result are the cosmological field equations:

$$\ddot{R} = -\frac{4\pi G}{3} R \sum_n \left(\rho_n + \frac{3p_n}{c^2}\right), \qquad (2)$$

$$\dot{R}^2 = \frac{8\pi G}{3} R^2 \sum_n \rho_n - kc^2. \qquad (3)$$

The cosmological field equations can be solved only if we know the content of the Universe: the number of different cosmological fluids and the corresponding functions $\rho_n$ and $p_n$. At this point, the relationship between physics and cosmology can be summarized in a single sentence addressed to physicists by cosmologists: *Please, tell us the content of the Universe and we will tell you how the Universe evolves in time*. The trouble is that the content of the Universe suggested by our best physics (i.e. the Standard Model of Particles and Fields) is apparently wrong. In order to explain observations, our best model of the Universe (*Inflationary ΛCDM* model) invokes mysterious content of the Universe (inflation field, dark matter and dark energy) and even with all this additional stuff of unknown nature we have the problem of the initial singularity (in Big-Bang theory), and we do not know the root of the cosmological constant problem and why matter dominates antimatter in the Universe.

### 3.1 A symmetric matter-antimatter Universe

According to our best knowledge we live in a Universe dominated by matter; this matter-antimatter asymmetry is considered one of the biggest mysteries in physics and cosmology. It is obvious that the gravitational repulsion between matter and antimatter cannot have any impact on the Universe if antimatter is not a significant part of the content of the Universe. The only, easily visible way, to "introduce" needed antimatter in the Universe is to abandon the current paradigm of matter-antimatter asymmetry and to assume *a symmetric Universe with equal amounts of matter and antimatter, with antimatter somehow hidden in the cosmic voids*. This common hypothesis is used in two different models named Dirac-Milne Cosmology[16-18] (a little bit misleading name because Dirac and Milne are not originators of this theory) and the Lattice Universe[19-21].

Before we continue the overview of these apparently "wild" models, let us underscore a crucial fact. Preliminary studies show a surprising and astonishing agreement with observations; more precisely models successfully pass the classical cosmological tests such as primordial nucleosynthesis, Type Ia supernovae and the Cosmic Microwave Background.

Of course, a Universe with huge and equal amounts of matter and antimatter is not empty, but it is *gravitationally empty* if $m_g + \bar{m}_g = 0$ i.e. if CPT symmetry is valid for gravity; total gravitational charge and average gravitational charge density are equal to zero. Consequently, Eq. (3) reduces to $\dot{R}^2 = -kc^2$; real solutions exist only for $k = -1$ (open Universe) and the scale factor of the Universe is a linear function of time, i.e. $R(t) = ct$. Hence, the FLRW metric (given by Eq. (1)) leads to the following metric for a symmetric matter-antimatter Universe:

$$ds^2 = c^2 dt^2 - c^2 t^2 \left[\frac{dr^2}{1+r^2} + r^2(d\theta^2 + sin^2\theta \, d\phi^2)\right]. \qquad (4)$$

By the way, metric (4) was used by Milne 8 decades ago, but of course without any involvement of antimatter; the point of Milne was that expansion exists without gravitation, i.e. in an empty Universe in the limit when General Relativity reduces to Special Relativity.

The hypothesis that antimatter is hidden in the cosmic voids (and there is apparently no other place to hide) is immediately in big trouble. Why antimatter in voids is invisible; why we see voids instead of stars, galaxies and clusters of galaxies made of antimatter?

As a possible explication of the invisibility of antimatter in voids, Dirac-Milne Cosmology[16-18] proposes what can be pictorially called "double antigravity", there is a first gravitational repulsion between matter and antimatter and a second *gravitational repulsion between antimatter and antimatter*. As a consequence of the second gravitational repulsion (i.e. repulsion between antimatter and antimatter) antimatter stars cannot exist (*this is of course a brutal violation of CPT symmetry*, but, as already noticed, not necessarily an argument against theory); hence, antimatter in voids is not in the form of stars and galaxies but just in the form of an invisible antimatter cloud that tends to expand (but expansion is limited by the gravitational repulsion of surrounding matter). However, in my opinion, this solution introduces a new problem: while the Universe remains gravitationally empty for matter it is not more gravitationally empty for antimatter (which is now repelled by both matter and antimatter) and apparently the scale factor of the Universe must be different for matter and antimatter.

As explained in a recent Physical Review article[18], a rigorous formulation of "double antigravity" is possible only within the framework of a bi-metric theory of gravity. In simple terms, *the postulated negative active gravitational charge of antimatter* would repel other antimatter only if the passive gravitational charge of antimatter is positive. On the other side the positive active gravitational charge of matter repels antimatter only if the passive gravitational charge of antimatter is negative. Hence, in one case the passive gravitational charge of antimatter must be positive and in the other case negative; in other words, gravitational properties of matter and antimatter cannot be described by simply assigning a combination of signs to the three types of Newtonian masses (i.e., the inertial mass, active gravitational mass and passive gravitational mass). Instead, Dirac-Milne Cosmology is formulated as a bi-metric theory. In brief, we witness a major reformulation of Dirac-Milne Cosmology, from a simple and elegant but insufficient metric (Eq. (4)) to a coherent bi-metric theory.

This is the right place for two comments. First, several authors have noted that our universe is very similar to a gravitationally empty or coasting universe[37,38] (neither accelerating nor decelerating); if it is true, Dirac-Milne Cosmology is a possible fundamental explanation. Second, there are two different $R(t) = ct$ cosmologies, Dirac-Milne Cosmology (a symmetric matter-antimatter Universe" and Melia-Shevchuk[39] "$R_H = ct$" Universe (a Universe without antimatter); apparently[5,18] a symmetric matter-antimatter Universe (and it is encouraging for Dirac-Milne Cosmology) is in better agreement with observations. Note that in a $R(t) = ct$ cosmology, $\dot{R}(t) = c$ and consequently the Hubble parameter H is exactly equal to $1/t$ ($H = \dot{R}/R = 1/t$).

Contrary to Dirac-Milne Cosmology, the Lattice Universe is based on strict respect of CPT symmetry. In fact, as a "prelude" to the Lattice Universe, Villata has argued[7] that *CPT and General Relativity are compatible only if matter and antimatter repel each other*. In the Lattice model, the Universe is considered to be like a gravitational lattice (analogous to an electrostatic lattice structure i.e. a crystal). The key result is that the alternation of the unlike (positive and negative) gravitational charges in the Universe *produces a net accelerated expansion* despite the equal amounts of the two components[20].

So far, scientists working on the development of Dirac-Milne Cosmology and the Lattice Universe haven't published any paper devoted to the study of a single galaxy. The gravitational field in a galaxy is much stronger that it can be according to the existing amount of Standard Model matter and our law of gravity; this anomalous gravitational field is usually attributed to dark matter, or a MOND-type modification of gravity.

It seems obvious (at least to the author of this Brief Review) that antimatter hidden in the voids cannot explain phenomenon of these mysterious central fields in galaxies. If so, Dirac-Milne Cosmology and the Lattice Universe are not a complete alternative to standard $\Lambda CDM$ Cosmology. One possible solution is to include dark matter in both Dirac-Milne Cosmology and the Lattice Universe; hence the content of the Universe would be equal amounts of matter and antimatter plus dark matter. However, the introduction of dark matter can violate the crucial assumption of a gravitationally empty Universe. Hence, in Dirac-Milne Cosmology and the Lattice Universe, the only acceptable dark matter must contain equal amounts of positive and negative gravitational mass; *dark matter must be dipolar*. It is amusing that dipolar dark matter, apparently needed in these two models, was proposed[30-34] and very successfully developed within the Dark Matter paradigm.

### 3.2 Quantum vacuum and virtual gravitational dipoles

There is a subtle way for a strong gravitational impact of antimatter in a Universe that is dominated by matter. Instead of antimatter hidden in the cosmic voids (which is the cornerstone of Dirac-Milne Cosmology and the Lattice Universe), the impact of antimatter can come from the quantum vacuum[22-

[29] which contains the same number of virtual particles and antiparticles. Historically, this is the first proposed paradigm, but we present it as the last one for pedagogical reasons.

It is well established that the quantum vacuum and matter immersed in it interact through electromagnetic, strong and weak interactions[40]. The open question is, if there are also gravitational interactions between the quantum vacuum and the immersed matter?

The heart of the paradigm is the following working hypothesis:

(**H**[1]) By their nature, quantum *vacuum fluctuations are virtual gravitational dipoles.*

By the way, the motivation for the hypothesis (**H**[1]) comes from the question what is the simplest possible solution to the cosmological constant problem. We know that the electric charge of the quantum vacuum is zero because virtual particles and antiparticles (which have the opposite electric charge and make an electric dipole) always appear in pairs. Consequently, it is obvious that the gravitational charge of the quantum vacuum would be zero (i.e. the quantum vacuum would be free of the cosmological constant problem) if particles and antiparticles have gravitational charge of the opposite sign. Of course, we don't know if this logically simplest solution and the real physical solution to the cosmological constant problem are the same.

According to the above hypothesis, a quantum vacuum fluctuation is a system of two gravitational charges (See a very schematic Figure 4) of the opposite sign; consequently, the total gravitational charge of a vacuum fluctuation is zero, but it has a non-zero gravitational dipole moment $\boldsymbol{p}_g$:

$$\boldsymbol{p}_g = m_g \boldsymbol{d}, |\boldsymbol{p}_g| < \frac{\hbar}{c}. \tag{5}$$

Here, $m_g$ denotes the magnitude of the gravitational charge, while, by definition, the vector $\boldsymbol{d}$ is directed from the antiparticle to the particle and has a magnitude d equal to the distance between them. The inequality in (5) follows from the fact that the size d of a quantum fluctuation is smaller than the reduced Compton wavelength (i.e. $d < \lambdabar_g = \hbar/m_g c$).

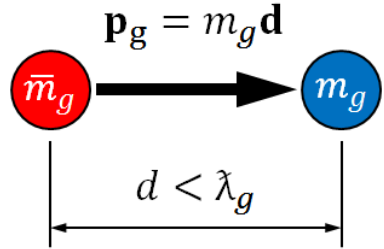

Figure 4. A virtual gravitational dipole is defined in analogy with an electric dipole: two gravitational charges of the opposite sign ($m_g > 0, m_g + \overline{m}_g = 0$) at a distance $d$ smaller than the corresponding reduced Compton wavelength $\lambdabar_g$. Note that the existence and impact of virtual electric dipoles is well established [40].

If gravitational dipoles exist, the gravitational polarization density $\boldsymbol{P}_g$, i.e. the gravitational dipole moment per unit volume, can be attributed to the quantum vacuum. It is obvious that the magnitude of the gravitational polarization density $\boldsymbol{P}_g$ satisfies the inequality $0 \leq |\boldsymbol{P}_g| \leq P_{gmax}$ where 0 corresponds to the *random orientations* of dipoles, while the maximal magnitude $P_{gmax}$ corresponds to the case of *saturation* (when all dipoles are aligned with the external field). The value $P_{gmax}$ must be a universal constant related to the gravitational properties of the quantum vacuum. Later we will discuss the possibility of the experimental determination of the eventual universal constant $P_{gmax}$.

If the external gravitational field is zero, the quantum vacuum may be considered like a fluid of randomly oriented gravitational dipoles (Figure 5a). In this case everything is equal to zero: the total gravitational charge, the gravitational charge density and the gravitational polarization density $\boldsymbol{P}_g$. Of course, such a vacuum is not a source of gravitation (note again that this is the simplest possible solution to the cosmological constant problem). However, the random orientation of virtual dipoles can be broken by the gravitational field of the immersed Standard Model matter. Massive bodies (particles, stars, planets, black holes…) but also many-body systems such as galaxies are surrounded by an invisible *halo* of the gravitationally polarized quantum vacuum, i.e. a region of non-random orientation of virtual gravitational dipoles (Figure 5b).

While the behaviour sketched in Figure 5.b is obvious for *permanent* gravitational dipoles, if you are not familiar with the quantum vacuum, you may wonder if it is also correct for extremely short-living *virtual* dipoles. Fortunately, the phenomenon of the electric polarization of the quantum vacuum[40] is well-established in Quantum Electrodynamics and this makes very plausible an analogous gravitational polarization of the quantum vacuum. More precisely, Quantum Electrodynamics (QED) is our first quantum field theory and the quantum vacuum (as an inherent part of QED) is one of the greatest discoveries in the history of science. One of important phenomena is the electric polarization of the quantum vacuum; in particular, the screening of an electric charge by the surrounding virtual electric dipoles. It is immediately obvious (to everyone familiar with the electric polarization of the quantum vacuum) that other kinds of polarization can exist as well. The gravitational polarization of the quantum vacuum is obvious if quantum vacuum fluctuations are virtual gravitational dipoles (defined in full analogy with electric dipoles).

The *spatial variation* of the gravitational polarization density generates[23,26,29] a *gravitational bound charge density* of the quantum vacuum

$$\rho_{qv} = -\boldsymbol{\nabla} \cdot \boldsymbol{P_g} \,. \tag{6}$$

You can consider this gravitational bound charge density to be an *effective gravitational charge density*, which acts as if there is a real non-zero gravitational charge. That is how the magic of polarization works; the *quantum vacuum is a source of gravity thanks to the immersed Standard Model matter*.

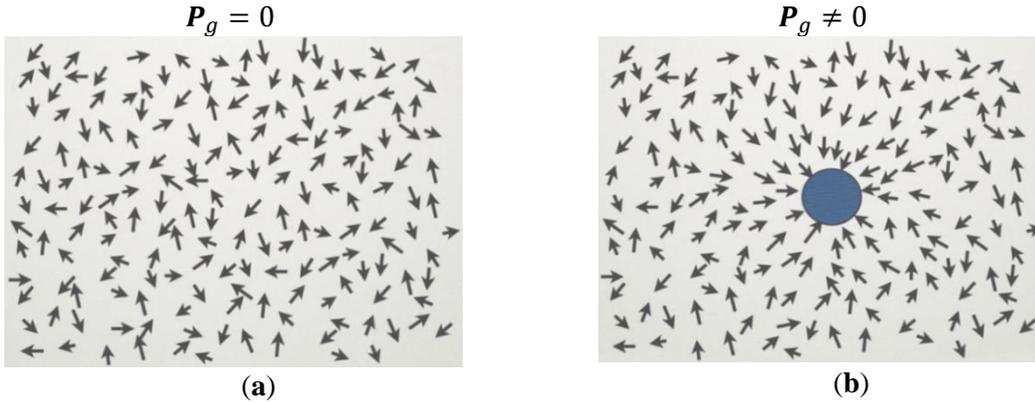

**Figure 5.** Schematic presentation of virtual gravitational dipoles in the quantum vacuum: (**a**) Randomly oriented gravitational dipoles (in the absence of an external field); (**b**) Halo of non-random oriented gravitational dipoles around a body with baryonic mass $M_b$.

The hypothesis ($H_1$) can be combined with a second hypothesis:

(**$H_2$**) Standard Model matter (i.e. matter made of quarks and leptons interacting through the exchange of gauge bosons) is the only content of the Universe.

The hypothesis ($H_2$) excludes dark matter and dark energy as the content of the Universe, while hypothesis ($H_1$) postulates the quantum vacuum as a cosmological fluid free of the cosmological constant problem. Together, these two hypotheses have the following series of intriguing consequences.

The phenomena usually attributed to hypothetical dark matter and dark energy can be considered[23,26,27,29] as a result of the gravitational polarization of the quantum vacuum by the immersed Standard Model matter. Locally, each halo of dark matter can be replaced by the halo of the polarized quantum vacuum[23,26,29]. Globally, all halos of the polarized quantum vacuum are a cosmological fluid, which, during the expansion of the Universe converts from a fluid with negative pressure - allowing an accelerated expansion of the Universe - to a fluid with zero pressure, which physically means the end of the accelerated expansion[29].

Additionally, the two hypotheses open the possibility that we live in a cyclic universe with cycles alternatively dominated by matter and antimatter[22,24,29]. The conversion of matter (or antimatter) of one cycle to antimatter (or matter) of the next cycle, happens in a cataclysmic event similar to the Big Bang, but at a significant macroscopic size (more precisely at a macroscopic size of the scale factor $R$ of the Universe); the cause of conversion is an extremely fast and tremendous creation of particle-antiparticle pairs from the quantum vacuum in an extremely strong gravitational field at a relatively small scale factor $R$ of the Universe. Consequently, at least mathematically, there is no initial singularity, there is no need for cosmic inflation and there is an elegant explanation of the matter-antimatter asymmetry in the universe: our universe is dominated by matter because the previous cycle was dominated by antimatter and the next one will be dominated by antimatter again.

### 3.3 Dipolar dark matter versus quantum vacuum

We have seen that two different paradigms, "*dipolar dark matter*"[30-34] and "*virtual gravitational dipoles*"[22-29] have a common concept: there is a cosmological fluid composed of *gravitational dipoles* and there is *gravitational* polarization[23,26,29] of this fluid (that is caused by immersed Standard Model Matter). Of course, this common concept leads to significant mathematical similarities; the basic Eq. (6) appears in both theories. However, despite common points, these two theories are fundamentally different and incompatible.

The first major difference is that in one theory *dark matter exists* while in the other theory *dark matter doesn't exist*. Dipolar Dark Matter is a successful unification of two apparently incompatible theories, MOND and Dark Matter. According to the Dipolar Dark Matter paradigm, *dark matter exists* but is composed of *permanent gravitational dipoles*; as a result of the gravitational polarization, such a kind of dark matter leads to MOND's equations. On the other hand, according to the paradigm that "*quantum vacuum fluctuations are virtual gravitational dipoles*" dark matter doesn't exist; *dark matter is just an illusion caused by the gravitational polarization of the quantum vacuum*[23,26,29].

It is illuminating to consider "*dipolar dark matter*" and "*virtual gravitational dipoles*" paradigms from the point of view of analogy between electric and gravitational polarization. Two well known cases of electric polarization are the polarization inside dielectric materials and the polarization inside the quantum vacuum[40]. Dipolar Dark Matter is the gravitational analogue of a dielectric; a still unknown (that remains to be discovered) kind of matter beyond the Standard Model of Particles and Fields. Virtual gravitational dipoles are the analogue of virtual electric dipoles in the quantum vacuum; there is no need to invoke new content of the Universe, Standard Model Matter and the quantum vacuum "enriched" with virtual gravitational dipoles can thus be the only content of the Universe.

Let us underscore the second fundamental difference. *The amount of dipolar dark matter in the Universe is a constant*; consequently, the ratio of dipolar dark matter and the baryonic matter in the Universe is a constant. Dipolar Dark Matter is a pressureless cosmological fluid. By contrast, *the effective gravitational charge of the quantum vacuum is not a constant* but depends on the scale radius of the Universe. Hence, the ratio of the effective gravitational charge of the quantum vacuum and baryonic matter is a variable; the polarized quantum vacuum is a cosmological fluid that in some periods of expansion can have negative pressure and consequently has the potential to explain phenomena usually attributed to dark energy.

The third crucial difference is the description of the gravitational field caused by an *isolated point-like body*. Within the Dipolar Dark Matter paradigm (and it is also valid in Dirac-Milne Cosmology and the Lattice Universe), the gravitational field of a point-like body (that is far from any other matter or dark matter) is classically described by Newton's inverse square law (or the Schwarzschild metric if General Relativistic effects cannot be neglected). However, within the quantum vacuum paradigm, a point-like body is not a point-like source of gravity, because it is *inseparable* from the halo of the polarized quantum vacuum around it; a halo that can extend to very large distances (for instance the halo of the Sun is much larger than the Solar System). Consequently, while there is no violation of Newton's law, the point-like body as one source of gravity, together with the *inevitable* effective gravitational charge of the polarized quantum vacuum as the second source of gravity, produce a composite gravitational field that cannot be described with the simple inverse square law.

In brief, the above differences tell us how crucial the nature of gravitational dipoles is; different kinds of gravitational dipoles can produce radically different effects.

# 4. Outlook and Astronomical observations

If, in experiments in our laboratories, antimatter falls upwards, it would be a scientific revolution, *but not confirmation of any of the proposed astrophysical and cosmological consequences*. It is obvious that eventual discovery of gravitational repulsion between matter and antimatter would not be confirmation of other hypothesis in the proposed theories; experiments with antihydrogen, muonium and positronium cannot tell us if there is antimatter hidden in voids, if there is gravitational repulsion between antimatter and antimatter, if quantum vacuum fluctuations behave as gravitational dipoles…

Astrophysical and cosmological phenomena can be revealed only by astronomical observations, and we are lucky that new generation telescopes (from the James Webb Space Telescope to the Extra-Large Telescope) will be operative nearly immediately after a surprising discovery that can come from CERN.

In general, significantly higher precision of astronomical measurements will increase our capacity to distinguish between predictions of different theories.

In addition to cosmological considerations it is illuminating to consider a black whole from the point of view of gravitational repulsion between matter and antimatter and three competing theories described in this review.

## 4.1 Black-white holes

Let us consider a black hole made of matter. If there is gravitational repulsion between matter and antimatter, such an object is a black hole for matter, but not a black hole for antimatter; any antimatter inside the horizon would be violently ejected by repulsion. We propose the name "black-white hole" for such entity, but please note that words black and white do not have the same meaning as in General Relativity.

Everyone thinks that experiments at CERN are just a measurement of the gravitational acceleration of antihydrogen. It is amusing that no one noticed that in fact, experiments at CERN are a test if *black-white holes* exist in the Universe. If antihydrogen falls upwards, black holes must be renamed to black-white holes; a black hole made from matter is a black hole for matter but a white hole for antimatter. If antihydrogen falls upwards it is an inevitable phenomenon independent of any theory.

Let us imagine matter falling into a black-white hole. A tiny fraction of falling matter will be ejected back in the form of high-energy antiparticles. Namely, as the result of the collisions of the infalling material (analogous to collisions in our accelerators), different kinds of antiparticles can be created inside the horizon and long-living antiparticles would be violently ejected outside the horizon. If black-white holes exist, they are an inevitable source of high energy positrons and antiprotons in cosmic rays.

An intriguing question is if two different signatures of these black-white holes have already been seen. The first signature may be an unexplained excess of high-energy positrons and antiprotons in cosmic rays[41] revealed by measurements with the Alpha Magnetic Spectrometer on the International Space Station. The second signature may be a recent detection, at the IceCube neutrino telescope at the South pole, of very high-energy (anti)neutrinos coming from the galactic centre[42]; apparently the Milky Way's supermassive black hole acts as mysterious "factory" of high-energy (anti)neutrinos.

There is a second, more subtle mechanism for creation of particle-antiparticle pairs deep inside the matter horizon. Let us remember that quantum vacuum is an inherent part[40] of the Standard Model of Particles and Fields and that under certain conditions virtual particle-antiparticle pairs from the quantum vacuum can be converted into real particles; we can create something from apparently nothing. For instance, an electron and a positron in a virtual pair can be converted to real ones in a sufficiently strong electric field accelerating them in the opposite direction. The same (i.e. creation of particle-antiparticle pairs from the quantum vacuum) can be done by the gravitational field if particles and antiparticles have gravitational charge of the opposite sign; the only difference is that the needed opposite acceleration is caused by a gravitational field.

Hence, black-white holes might radiate because of particle-antiparticle creation from the quantum vacuum[25]. It is obvious that this is a model dependant mechanism (for instance not valid in Dirac-Milne Cosmology and the Lattice Universe). A major question is if Hawking radiation can coexist with quantum vacuum radiation? The answer is: No. Hawking radiation depends on the heretofore assumed model of the gravitational properties of the quantum vacuum. Hawking calculations correspond to the case of gravitational monopoles and cannot be valid if the quantum vacuum is composed of gravitational dipoles.

### 4.2 Trans-Neptunian binaries and quantum vacuum

As a model-dependent phenomenon (not existing in Dirac-Milne Cosmology and the Lattice Universe) let us mention that the quantum vacuum might have a tiny impact on orbits of celestial bodies in the Solar System; it is the gravitational analogue of the Lamb shift[40] (i.e. of the impact of the quantum vacuum on energy levels of electrons in atoms) in Quantum Electrodynamics. Apparently, the most promising way to reveal such an impact of the quantum vacuum is the study of orbits of tiny satellites in some trans-Neptunian binaries[14-15]. The key point is that in an *isolated binary*, the quantum vacuum causes a perihelion shift per orbit $\Delta\omega_{qv}$ which is directly proportional to the maximal magnitude $P_{gmax}$ of the gravitational polarization density $\boldsymbol{P}_g$ (See section 3.2 after Eq. (5)).

### 4.3 Brief summary

Our current understanding of the Universe ($\Lambda CDM$ Cosmology) is both, a fascinating intellectual achievement and the source of *the greatest crisis in the history of physics*. We do not know the nature of what we call an inflation field, dark matter and dark energy; we do not know why matter dominates antimatter in the Universe and what the root of the cosmological constant problem is.

In about three years from now experiments will end the already 6 decades old theoretical dispute of whether antimatter falls down or falls up. Experiments will tell us if antimatter gravity is crucial or not in the understanding of the greatest mysteries of contemporary physics, astrophysics and cosmology.

So far there are three pioneering theories (Dirac-Milne Cosmology, the Lattice Universe and Cosmology with quantum vacuum fluctuations as virtual gravitational dipoles) that anticipate repulsive gravity as the outcome of the forthcoming experiments.

**Table 1**. Cosmological models based on the gravitational repulsion between matter and antimatter – similarities, differences and comparison with the inflation-based $\Lambda CDM$ Cosmology

|  | **Dirac-Milne Cosmology** | **The Lattice Universe** | **Hajdukovic's Cosmology** |
|---|---|---|---|
| Description of gravitational interactions | Matter attracts matter | | |
| | Matter and antimatter repel each other | | |
| | Antimatter repels antimatter (CPT is violated) | Antimatter attracts antimatter (CPT Symmetry is not violated) | |
| Matter-antimatter content of the Universe | Matter-antimatter symmetric Universe (i.e. Universe contains the same amounts of matter and antimatter with antimatter "hidden" in the voids) | | Our cycle of the Universe is dominated by matter |
| Quantum vacuum | So far, these models do not include quantum vacuum as a source of gravity | | Quantum vacuum is a crucial source of gravity |
| Matter-energy content of the Universe | In current versions of models, the only content of the Universe are equal amounts of matter and antimatter. Apparently, in the forthcoming improved versions of these models a kind of dark matter (or something replacing dark matter) must be included. | | The Standard Model Matter and quantum vacuum composed of virtual gravitational dipoles |
| Cornerstones of $\Lambda CDM$ | | | |
| Dark matter | Phenomena usually attributed to Dark matter remain to be explained in both models | | What we call DM and DE are local and global effects of gravitational polarization of the quantum vacuum |
| Dark energy | There is no need for Dark energy | | |
| Inflation field | Apparently, in all 3 models there is no need for the inflation field | | |
| Cosmological Constant Problem | So far, this problem is not considered in these models | | A solution is proposed |

As explained in this Review (and summarized in Table 1) these theories are radically different and mutually excluding which is good for the development of science. For the first time we have a theory that (free of the cosmological constant problem) considers the quantum vacuum as an inevitable source of gravity and provides a common explanation of phenomena usually attributed to dark matter and dark energy; without the need for cosmic inflation and replete with the striking explanation of matter-antimatter asymmetry in the Universe (a cyclic universe alternatively dominated by matter and antimatter). The other two models (Dirac-Milne Cosmology and the Lattice Universe) *do not use the quantum vacuum as a source of gravity* but they achieved intriguing initial success assuming a symmetric matter-antimatter Universe.

Let us wait and see if *antimatter gravity experiments* will be the birth of a new scientific revolution.